\documentclass[twocolumn,showpacs,preprintnumbers,prl]{revtex4}
\usepackage{epsfig,amsmath,amssymb}

\newcommand{\beq}{\begin{equation}}
\newcommand{\eeq}{\end{equation}}
\newcommand{\bea}{\begin{eqnarray}}
\newcommand{\eea}{\end{eqnarray}}
\newcommand{\Fig}[1]{Fig.\,\ref{#1}}

\newcommand{\Eq}[1]{Eq.\,(\ref{#1})}
\newcommand{\eq}[1]{(\ref{#1})}
\newcommand{\Eqs}[1]{Eqs.\,(\ref{#1})}
\newcommand{\Eqsand}[2]{Eqs.\,(\ref{#1}) and (\ref{#2})}
\newcommand{\Eqsto}[2]{Eqs.\,(\ref{#1})--(\ref{#2})}

\newcommand{\Tab}[1]{Table\,\ref{#1}}
\newcommand{\f}{\frac}
\newcommand{\non}{\nonumber}

\newcommand{\as}{\alpha_s}
\newcommand{\aem}{\alpha}
\newcommand{\aemMSbar}{\alpha_{\overline{\rm MS}}}
\newcommand{\sws}{\sin^2 \theta_{\scriptscriptstyle W}}
\newcommand{\swsMSbar}{\sin^2 \hat{\theta}_{\scriptscriptstyle
W}^{\overline{\rm MS}}}    
\newcommand{\swq}{\sin^4 \theta_{\scriptscriptstyle W}}
\newcommand{\MW}{M_{\scriptscriptstyle W}}
\newcommand{\MZ}{M_{\scriptscriptstyle Z}}
\newcommand{\mc}{m_c}
\newcommand{\mb}{m_b}
\newcommand{\mt}{m_t}
\newcommand{\mtpole}{M_t}

\newcommand{\muh}{\mu_{\scriptscriptstyle W}}

\newcommand{\mut}{\mu_t}
\newcommand{\muc}{\mu_c}
\newcommand{\mub}{\mu_b}
\newcommand{\cf}{C_{\scriptstyle F}}
\newcommand{\ca}{C_{\scriptstyle A}}

\newcommand{\nf}{f}

\newcommand{\BR}{{\cal{B}}}
\newcommand{\GeV}{{\rm \ GeV}}

\newcommand{\MSbar}{\overline{\rm MS}}
\newcommand{\re}{{\rm Re}}

\newcommand{\ord}{{\cal O}}

\newcommand{\Ktopinunu}{K^+ \to \pi^+ \nu \bar{\nu}}

\newcommand{\Ktomumu}{K_L \to \mu^+ \mu^-}
\newcommand{\Ktogammagamma}{K_L \to \gamma \gamma}
\newcommand{\Ktoggmm}{K_L \to \gamma^\ast \gamma^\ast \to \mu^+ \mu^-}

\newcommand{\Ktomunu}{K^+ \to \mu^+ \nu_\mu}

\newcommand{\Pc}{P_c}
\newcommand{\ckm}{{\tt CKMFITTER}}

\begin{document}


\preprint{TTP06-16; ZU-TH 10/06; hep-ph/0605203}

\title{
\boldmath
Charm-Quark Contribution to $\Ktomumu$ at Next-to-Next-to-Leading
Order       
\unboldmath}
\author{Martin Gorbahn${}^1$ and Ulrich Haisch${}^2$}   

\affiliation{
$^1\!\!\!$ Institut f\"ur Theoretische Teilchenphysik, Universit\"at 
Karlsruhe, D-76128 Karlsruhe, Germany \\
$^2\!\!\!$ Institut f\"ur Theoretische Physik, Universit\"at Z\"urich,
CH-8057 Z\"urich, Switzerland
}

\date{September 25, 2006}

\begin{abstract}
\noindent
We calculate the charm-quark contribution to the decay $\Ktomumu$ in
next-to-next-to-leading order of QCD. This new contribution reduces
the theoretical uncertainty in the relevant parameter $\Pc$ from $\pm
22 \%$ down to $\pm 7 \%$, corresponding to scale uncertainties of
$\pm 3 \%$ and $\pm 6 \%$ in the short-distance part of the branching
ratio and the determination of the Wolfenstein parameter $\bar{\rho}$
from $\Ktomumu$.  The error in $\Pc = 0.115 \pm 0.018$ is now in equal
shares due to the combined scale uncertainties and the current
uncertainty in the charm-quark mass. We find $\BR (\Ktomumu)_{\rm SD}
= \left ( 0.79 \pm 0.12 \right ) \times 10^{-9}$, with the present
uncertainty in the Cabibbo-Kobayashi-Maskawa matrix element $V_{td}$
being the dominant individual source in the quoted error.
\end{abstract}

\pacs{12.15.Hh, 12.38.Bx, 13.20.Eb}

\maketitle

The study of the rare process $\Ktomumu$ has played a central role in   
unraveling the flavor content and structure of the standard model (SM)
of electroweak interactions \cite{history}. These glory days have
passed, but still today $\Ktomumu$ provides useful information on the
short-distance dynamics of $| \Delta S | = 1$ 
flavor-changing-neutral-current transitions despite the
fact that its decay amplitude is dominated by the long-distance two
photon contribution $\Ktoggmm$. While the absorptive part of the
latter correction is calculable with high precision in terms of the
$\Ktogammagamma$ rate the corresponding dispersive part represents a
significant source of theoretical uncertainty. In fact long- and
short-distance dispersive pieces cancel against each other in
large parts and the measured total $\Ktomumu$ rate
\cite{Ambrose:2000gj} is nearly saturated by the absorptive two photon
contribution. The precision in the determination of the dispersive
pieces therefore controls the accuracy of possible bounds on the real
part of the Cabibbo-Kobayashi-Maskawa (CKM) element $V_{td}$ or,
equivalently, the Wolfenstein parameter $\bar{\rho}$. In view of the
recent experimental \cite{experiment} and theoretical
\cite{Isidori:2003ts} developments concerning the dispersive
long-distance part of the $\Ktomumu$ decay amplitude it is also
worthwhile to improve the theoretical accuracy of the associated
short-distance contribution. This is the purpose of this Letter.  

The branching ratio for the dispersive short-distance part of
$\Ktomumu$ can be written as \cite{Buchalla:1993wq}   
\begin{widetext}
\begin{gather} 
\label{eq:BR}
\BR ( \Ktomumu )_{\rm SD} = \kappa_\mu \left [ \f{\re
\lambda_t}{\lambda^5} Y(x_t) + \f{\re \lambda_c}{\lambda} \Pc \right
]^2 \, , \\ \label{eq:kappamu} \kappa_\mu \equiv \f{\aem^2 \, \BR
(\Ktomunu)}{\pi^2 \swq} \f{\tau (K_L)}{\tau (K^+)} \lambda^8 = \left (
2.009 \pm 0.017\right ) \times 10^{-9} \left ( \f{\lambda}{0.225}
\right )^8 \, ,   
\end{gather}
\end{widetext}
where $\lambda_i \equiv V^\ast_{is} V_{id}$ denote the relevant CKM
factors. There is also a short-distance two-loop electroweak
contribution in the two-photon mediated decay amplitude
\cite{Eeg:1996pr}.  Following \cite{Isidori:2003ts}, where this
contribution is included in the two-photon correction itself, we do
not add it to the short-distance contribution in \Eq{eq:BR}. The
apparent strong dependence of $\BR (\Ktomumu)_{\rm SD}$ on $\lambda
\equiv | V_{us} |$ is spurious as $\Pc$ is proportional to
$1/\lambda^4$. In quoting the value for $\Pc$ we will set $\lambda =
0.225$. The electromagnetic coupling $\aem$ and the weak mixing angle
$\sws$ entering $\BR (\Ktomumu)$ are naturally evaluated at the
electroweak scale \cite{Bobeth:2003at}. Then the leading term in the
heavy top expansion of the electroweak two-loop corrections to $Y
(x_t)$ amounts to typically $-1.5 \%$ for the modified minimal
subtraction scheme ($\MSbar$) definition of $\aem$ and $\sws$
\cite{Buchalla:1997kz}. In obtaining the numerical value of
\Eq{eq:kappamu} we have employed $\aem \equiv \aemMSbar (\MZ) =
1/127.9$, $\sws \equiv \swsMSbar = 0.231$, and $\BR (\Ktomunu) = \left
  ( 63.39 \pm 0.18 \right ) \times 10^{-2}$ \cite{PDG}.

The function $Y(x_t)$ in \Eq{eq:BR} depends on the top quark $\MSbar$
mass through $x_t \equiv \mt^2 (\mut)/\MW^2$. It originates from
$Z$-penguin and electroweak box diagrams with an internal top
quark. As the relevant operator has a vanishing anomalous dimension
and the energy scales involved are of the order of the electroweak
scale or higher, the function $Y(x_t)$ can be calculated within
ordinary perturbation theory. It is known through next-to-leading
order (NLO) \cite{Y, Buchalla:1998ba}, with a scale uncertainty due to
the top quark matching scale $\mut = \ord (\mt)$ of slightly more than
$\pm 2 \%$. Converting the top quark pole mass of $\mtpole = \left (
  172.5 \pm 2.3 \right ) \!\! \GeV$ \cite{Group:2006qt} at three loops
to $\mt (\mtpole)$ \cite{pole} and relating $\mt (\mtpole)$ to $\mt
(\mt) = \left ( 162.8 \pm 2.2 \right ) \!\! \GeV$ using the one-loop
renormalization group (RG), we find $Y (x_t) = 0.950 \pm 0.049$. The
given uncertainty combines linearly an error of $\pm 0.029$ due to the
error of $\mt (\mt)$ and an error of $\pm 0.020$ obtained by varying
$\mut$ in the range $60 \GeV \le \mut \le 240 \GeV$.

The calculable parameter $\Pc$ entering \Eq{eq:BR} results from
$Z$-penguin and electroweak box diagrams involving internal
charm-quark exchange. As now both high- and low-energy scales, namely,
$\muh = \ord (\MW)$ and $\muc = \ord (\mc)$, are involved, a complete
RG analysis of this term is required. In this manner, large logarithms
$\ln (\muh^2/\muc^2)$ are resummed to all orders in $\as$. The large
scale uncertainty due to $\muc$ of $\pm 44 \%$ in the leading order
result was a strong motivation for the NLO analysis of this
contribution \cite{Buchalla:1993wq, Buchalla:1998ba}.

\begin{table}[!t]
\caption{\sf Input parameters used in the numerical analysis of $\Pc$,
$\BR (\Ktomumu)_{\rm SD}$, and $\bar{\rho}$.}  
\vspace{1mm}
\begin{tabular}{l@{\hspace{5mm}}c@{\hspace{5mm}}c}
\hline \hline
Parameter & Value $\pm$ Error & Reference \\[0.25mm]
\hline
&&\\[-3.5mm]
$\mc (\mc)$ \hspace{0.2mm} [GeV] & $1.30 \pm 0.05$ & \cite{charm}, our
average \\     
$\as (\MZ)$ & $0.1187 \pm 0.0020$ & \cite{PDG} \\[0.5mm]
$\re \lambda_t$ \hspace{0.25mm} $[10^{-4}]$ & $-3.11^{+0.13}_{-0.14}$ 
& \cite{Charles:2004jd} \\[0.5mm] 
$\re \lambda_c$ & $-0.22098^{+0.00095}_{-0.00091}$ &
\cite{Charles:2004jd} \\[1mm]   
\hline \hline
\end{tabular}
\label{tab:input}
\end{table}

Performing the RG running from $\muh$ down to $\mub = \ord (\mb)$ in
an effective five-flavor theory and the subsequent evolution from
$\mub$ down to $\muc$ in an effective four-flavor theory, we obtain at
NLO          
\beq \label{eq:PcNLO}
\begin{split}
\Pc & = 0.106 \pm 0.023_{\rm theor} \pm 0.009_{\mc} \pm
0.001_{\as} \\
& = \left ( 0.106 \pm 0.034 \right ) \left ( \f{0.225}{\lambda} \right
)^4 \, ,  
\end{split}
\eeq 
where the parametric errors correspond to the ranges of the
charm-quark $\MSbar$ mass $\mc (\mc)$ and the strong coupling constant
$\as (\MZ)$ given in \Tab{tab:input}. The final error has been
obtained by performing a detailed analysis of the individual sources
of uncertainty entering the NLO prediction of $\Pc$ using a modified
version of the \ckm \hspace{0.4mm} package \cite{Charles:2004jd}. The
same statistical treatment of errors will be applied in
\Eqs{eq:BRNLO}, \eq{eq:PcNNLO}, and \eq{eq:BRNNLO}.

\begin{figure}[!t]
\begin{center}
\scalebox{0.7}{
\begingroup%
  \makeatletter%
  \newcommand{\GNUPLOTspecial}{%
    \@sanitize\catcode`\%=14\relax\special}%
  \setlength{\unitlength}{0.1bp}%
\begin{picture}(3600,2160)(0,0)%
\special{psfile=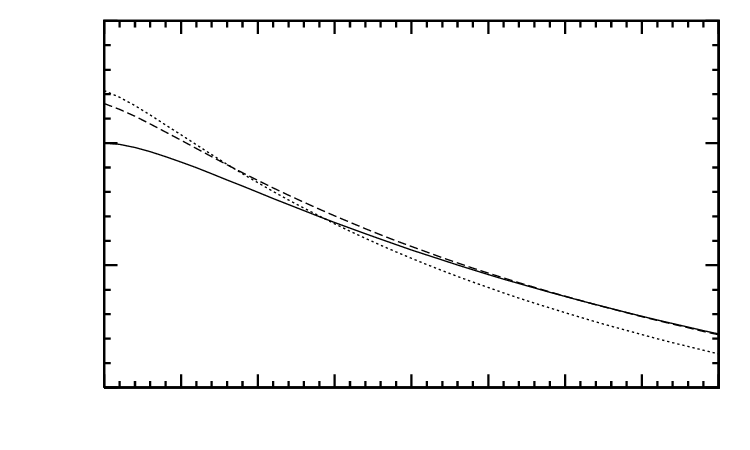 llx=0 lly=0 urx=360 ury=216 rwi=3600}
\put(1975,50){\makebox(0,0){\scalebox{1.3}{$\muc$ [GeV]}}}%
\put(100,1180){%
\special{ps: gsave currentpoint currentpoint translate
270 rotate neg exch neg exch translate}%
\makebox(0,0)[b]{\shortstack{\scalebox{1.3}{$\Pc$}}}%
\special{ps: currentpoint grestore moveto}%
}%
\put(3450,200){\makebox(0,0){ 3}}%
\put(3081,200){\makebox(0,0){ 2.75}}%
\put(2713,200){\makebox(0,0){ 2.5}}%
\put(2344,200){\makebox(0,0){ 2.25}}%
\put(1975,200){\makebox(0,0){ 2}}%
\put(1606,200){\makebox(0,0){ 1.75}}%
\put(1238,200){\makebox(0,0){ 1.5}}%
\put(869,200){\makebox(0,0){ 1.25}}%
\put(500,200){\makebox(0,0){ 1}}%
\put(450,2060){\makebox(0,0)[r]{ 0.14}}%
\put(450,1473){\makebox(0,0)[r]{ 0.12}}%
\put(450,887){\makebox(0,0)[r]{ 0.1}}%
\put(450,300){\makebox(0,0)[r]{ 0.08}}%
\end{picture}%
\endgroup
 }\\[1mm]
\scalebox{0.7}{
\begingroup%
  \makeatletter%
  \newcommand{\GNUPLOTspecial}{%
    \@sanitize\catcode`\%=14\relax\special}%
  \setlength{\unitlength}{0.1bp}%
\begin{picture}(3600,2160)(0,0)%
\special{psfile=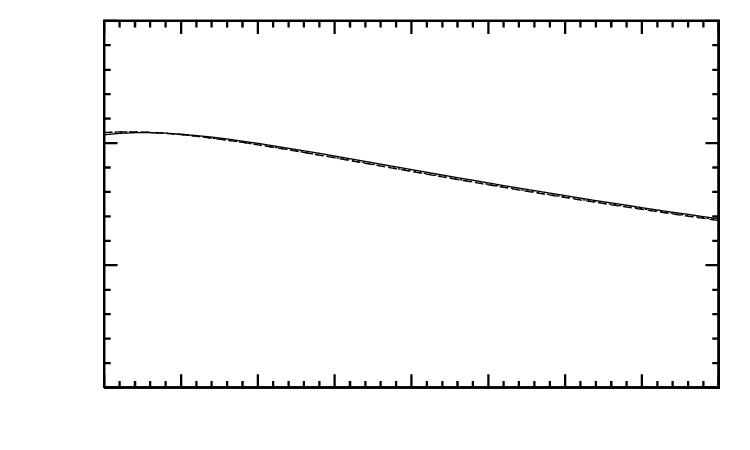 llx=0 lly=0 urx=360 ury=216 rwi=3600}
\put(1975,50){\makebox(0,0){\scalebox{1.3}{$\muc$ [GeV]}}}%
\put(100,1180){%
\special{ps: gsave currentpoint currentpoint translate
270 rotate neg exch neg exch translate}%
\makebox(0,0)[b]{\shortstack{\scalebox{1.3}{$\Pc$}}}%
\special{ps: currentpoint grestore moveto}%
}%
\put(3450,200){\makebox(0,0){ 3}}%
\put(3081,200){\makebox(0,0){ 2.75}}%
\put(2713,200){\makebox(0,0){ 2.5}}%
\put(2344,200){\makebox(0,0){ 2.25}}%
\put(1975,200){\makebox(0,0){ 2}}%
\put(1606,200){\makebox(0,0){ 1.75}}%
\put(1238,200){\makebox(0,0){ 1.5}}%
\put(869,200){\makebox(0,0){ 1.25}}%
\put(500,200){\makebox(0,0){ 1}}%
\put(450,2060){\makebox(0,0)[r]{ 0.14}}%
\put(450,1473){\makebox(0,0)[r]{ 0.12}}%
\put(450,887){\makebox(0,0)[r]{ 0.1}}%
\put(450,300){\makebox(0,0)[r]{ 0.08}}%
\end{picture}%
\endgroup
 }
\end{center}
\vspace{-5mm}
\caption{\sf $\Pc$ as a function of $\muc$ at NLO (upper plot) and
NNLO (lower plot). The three different lines correspond to three
different methods of computing $\as (\muc)$ from $\as (\MZ)$ (see
text).}        
\label{fig:plots}
\end{figure}
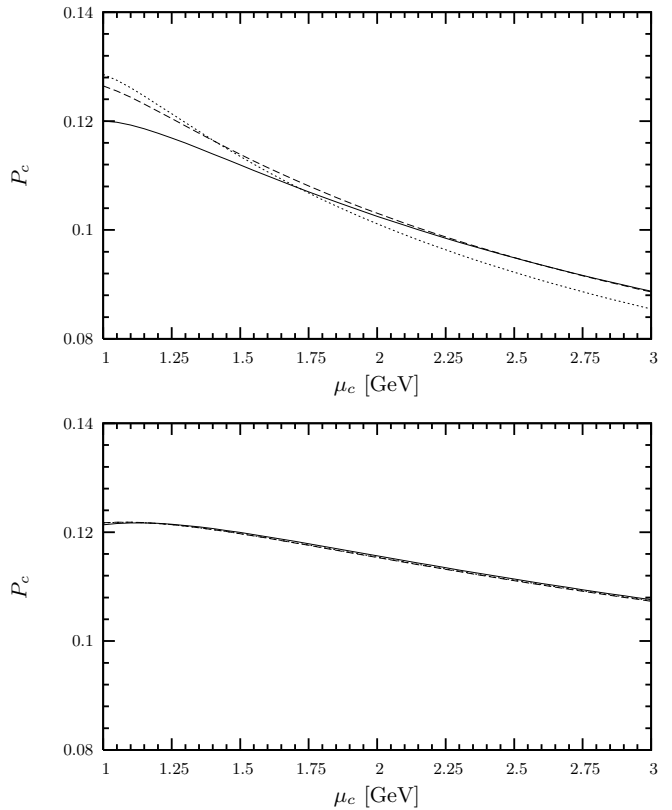

The dependence of $\Pc$ on $\muc$ can be seen in \Fig{fig:plots}. The 
solid line in the upper plot shows the NLO result obtained by 
evaluating $\as (\muc)$ from $\as (\MZ)$ solving the RG equation of
$\as$ numerically, while the dashed and dotted lines are obtained by
first determining the scale parameter $\Lambda_{\MSbar}$ from $\as
(\MZ)$, either using the explicit solution of the RG equation of $\as$
or by solving the RG equation of $\as$ iteratively for
$\Lambda_{\MSbar}$, and subsequently calculating $\as (\muc)$ from 
$\Lambda_{\MSbar}$. The corresponding two-loop values for $\as (\muc)$
have been obtained with the program {\tt RUNDEC}
\cite{Chetyrkin:2000yt}. Obviously, the difference between the 
three curves is due to higher order terms and has to be regarded as
part of the theoretical error. With its size of $\pm 0.006$ it is
almost comparable to the variation of the NLO result due to $\muc$,
amounting to $\pm 0.016$. In \cite{Buchalla:1993wq} a larger value for
the latter uncertainty has been quoted. The observed difference is
related to the definition of the charm-quark mass. Replacing $\mc
(\mc)$ in the logarithms $\ln ( \muc^2/\mc^2 )$ of the one-loop matrix
elements by the more appropriate $\mc (\muc)$ leads to a significant
reduction of the dependence of $\Pc$ on $\muc$. A detailed discussion
of this issue can be found in \cite{Buras:2006gb}. Finally, while in
\cite{Buchalla:1993wq} only $\muc$ was varied, the theoretical error
given in \Eq{eq:PcNLO} includes also the dependence on $\mub$ and
$\muh$ of combined $\pm 0.001$. The specified scale uncertainties
correspond to the ranges $1 \GeV \le \muc \le 3 \GeV$, $2.5 \GeV \le
\mub \le 10 \GeV$, and $40 \GeV \le \muh \le 160 \GeV$.  

Using the input parameters listed in \Tab{tab:input}, we find from 
\Eqsto{eq:BR}{eq:PcNLO} at NLO      
\begin{align} \label{eq:BRNLO}
\BR (\Ktomumu)_{\rm SD} & = \left ( 0.77 \pm 0.08_{\Pc} \pm 0.08_{\rm
other} \right ) \times 10^{-9} \non \\
& = \left ( 0.77 \pm 0.16 \right ) \times 10^{-9} \, ,   
\end{align}
where the second error in the first line collects the uncertainties
due to $\kappa_\mu$, $Y (x_t)$, and the CKM elements.   

\begin{figure}[!t]
\vspace{10mm}
\begin{center}
\scalebox{0.5}{\includegraphics{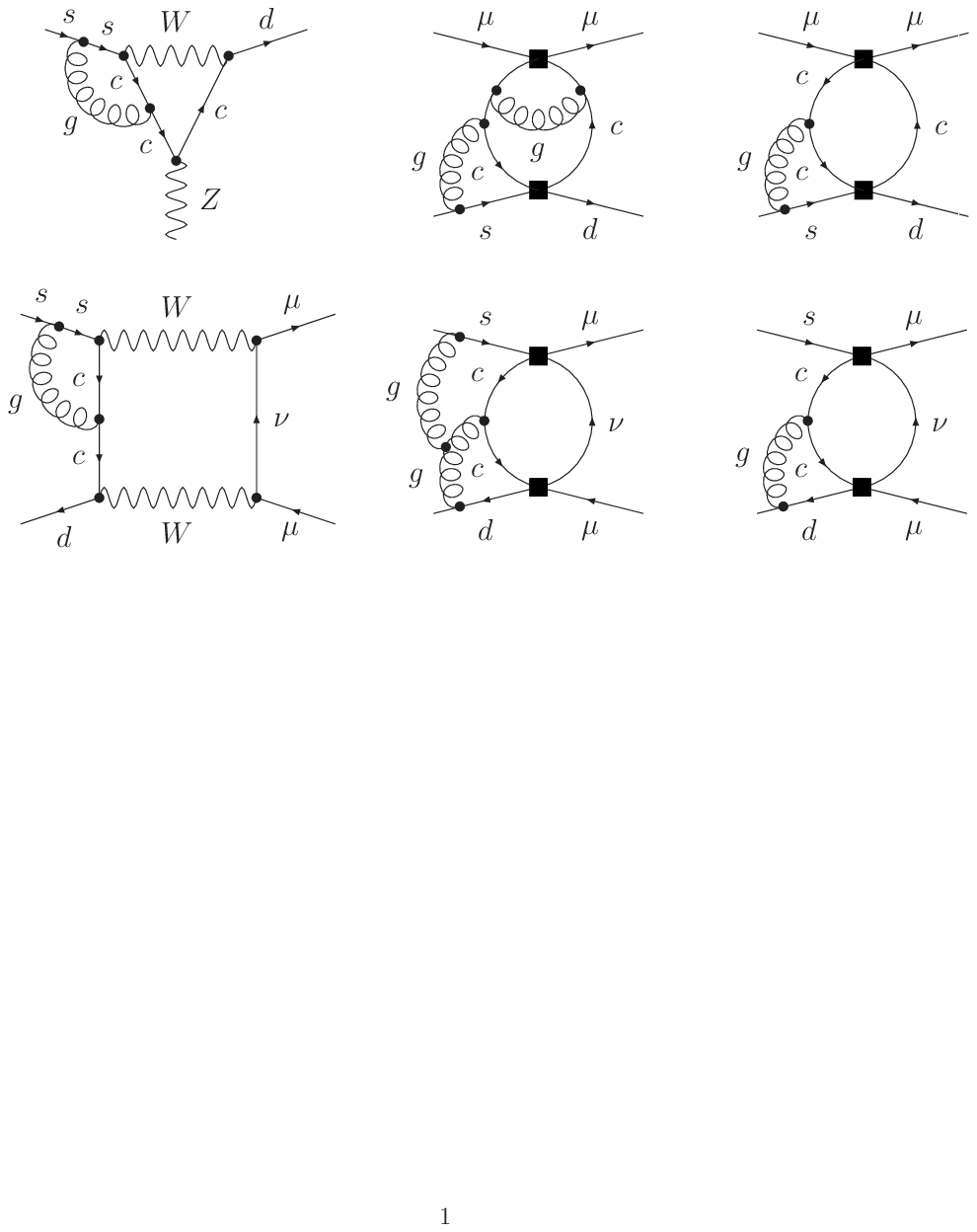}}
\end{center}
\vspace{-15mm}
\caption{Examples of Feynman diagrams arising in the full SM (left
column), describing the mixing of operators (center column) and the
matrix elements (right column) in the $Z$-penguin (upper row) and the
electroweak box (lower row) sector. Only the divergent pieces of the
diagrams displayed in the center column have to be computed, while the
Feynman graphs shown on the left- and right-hand side are needed
including their finite parts.}        
\label{fig:diagrams}
\end{figure}

As the uncertainties in \Eqsand{eq:PcNLO}{eq:BRNLO} coming from
$\mtpole$, $\mc (\mc)$ and the CKM parameters should be decreased in 
the coming years it is also desirable to reduce the theoretical
uncertainty in $\Pc$. To this end, we here extend the NLO analysis of
$\Pc$ presented in \cite{Buchalla:1993wq, Buchalla:1998ba} to the
next-to-next-to-leading order (NNLO). This requires the computation of
three-loop anomalous dimensions of certain operators and of certain
two-loop contributions. 

The main components of the NNLO calculation, which aims at resumming
all $\ord (\as^n \ln^{n-1} ( \muh^2/\muc^2 ))$ logarithms in $\Pc$,
are $(i)$ the $\ord (\as^2)$ matching corrections to the relevant
Wilson coefficients arising at $\muh$, $(ii)$ the $\ord (\as^3)$
anomalous dimensions describing the mixing of the dimension-six and
-eight operators, $(iii)$ the $\ord (\as^2)$ threshold corrections to
the Wilson coefficients originating at $\mub$, and $(iv)$ the $\ord
(\as^2)$ matrix elements of some of the operators emerging at
$\muc$. To determine the contributions of type $(i)$, $(iii)$, and
$(iv)$ one must calculate two-loop Green functions in the full SM and
in effective theories with five or four flavors. Sample diagrams for
steps $(i)$ and $(iv)$ are shown in the left and right columns of
\Fig{fig:diagrams}. The contributions $(ii)$ are found by calculating
three-loop Green functions with operator insertions. Sample diagrams
with a double insertion of dimension-six operators are shown in the
center column of \Fig{fig:diagrams}. 

The $Z$-penguin contribution can be trivially obtained from that in
$\Ktopinunu$, which has been recently computed at NNLO
\cite{Buras:2006gb, Buras:2005gr}. The electroweak box contribution on
the other hand is slightly different for $\Ktomumu$ and $\Ktopinunu$
since the lepton line in the corresponding Feynman diagrams is
reversed and thus requires a new calculation. A comprehensive
discussion of the technical details of the matching and the
renormalization of the effective theory can be found in
\cite{Buras:2006gb}.  

In the following we present only the final result for the $\ord
(\as^2)$ matching correction $C_\mu^{B (2)}$, the $\ord (\as^3)$
anomalous dimension $\gamma_\mu^{B (2)}$, and the $\ord (\as^2)$
matrix element $r_\mu^{B (2)}$. Employing the operator basis of
\cite{Buchalla:1993wq, Buchalla:1998ba} we obtain for the standard
choices of Casimir operators $\ca = 3$, $\cf = 4/3$, and $\nf$ active
quark flavors 
\beq \label{eq:nnlopieces}
\begin{split}
C_\mu^{B (2)} & = \f{416}{3} + \f{16 \hspace{0.4mm} \pi^2}{3} +
\f{272}{3} \ln \f{\muh^2}{\MW^2} + 16 \ln^2 \f{\muh^2}{\MW^2} \, ,
\\[1mm] 
\gamma_\mu^{B (2)} & = \f{27032}{9} - 1088 \hspace{0.6mm} \zeta (3)
-\f{1040}{9} \nf \, 
, \\[1mm] 
r_\mu^{B (2)} & = -\f{112}{3} -\f{80}{3} \ln \f{\muc^2}{\mc^2} - 16
\ln^2 \f{\muc^2}{\mc^2} \, . 
\end{split} 
\eeq 
Here $\zeta (x)$ is the Riemann zeta function with the value
$\zeta (3) \approx 1.20206$ and $\mc \equiv \mc (\muc)$ denotes the
charm-quark $\MSbar$ mass. Our results for the NLO Wilson coefficient,
the anomalous dimension and the matrix element agree with the findings
of \cite{Buchalla:1998ba} where an error made in the original
calculation \cite{Buchalla:1993wq} has been corrected.

\begin{table}[!t]
\caption{\sf The coefficients $\kappa_{ijkl}$ arising in the approximate
formula for $\Pc$ at NNLO.} 
\vspace{-3mm}
\begin{center}
\begin{tabular}{lll}
\hline \hline 
$\kappa_{1000} = -0.5373$ & $\kappa_{0100} = -6.0472$ &
$\kappa_{0010} = -0.0956$ \\
$\kappa_{0001} = 0.0114$ & $\kappa_{1100} = 3.9957$ & 
$\kappa_{1010} = 0.3604$ \\ 
$\kappa_{0110} = 0.0516$ & $\kappa_{0101} = -0.0658$ &  
$\kappa_{2000} = -0.1767$ \\ 
$\kappa_{0200} = 16.4465$ & $\kappa_{0020} = -0.1294$ & 
$\kappa_{0030} = 0.0725$ \\  
\hline \hline
\end{tabular}              
\label{tab:kappas}
\end{center}
\end{table}

The analytic expression for $\Pc$ including the complete NNLO
corrections is too complicated and too long to be presented
here. Instead setting $\lambda = 0.225$, $\mt (\mt) = 162.8 \! \GeV$
and $\muh = 80.0 \! \GeV$ we derive an approximate formula for $\Pc$
that summarizes the dominant parametric and theoretical uncertainties
due to $\mc (\mc)$, $\as (\MZ)$, $\muc$, and $\mub$. It reads
\beq \label{eq:masterformula}
\begin{split}
\Pc & = 0.1198 \left ( \f{\mc (\mc)}{1.30 \! \GeV} \right )^{2.3595}
\left ( \f{\as (\MZ)}{0.1187} \right )^{6.6055} \\[1mm]
& \times \left ( 1 + \sum_{i,j,k,l} \kappa_{ijlm} L_{\mc}^i L_{\as}^j
L_{\muc}^k L_{\mub}^l \right ) ,
\end{split}       
\eeq
where 
\beq \label{eq:defLs}
\begin{split}
\begin{aligned} 
L_{\mc} & = \ln \left ( \f{\mc (\mc)}{1.30 \! \GeV} \right ) \, , & 
L_{\as} & = \ln \left ( \f{\as (\MZ)}{0.1187} \right ) \,
, \\[1mm]
L_{\muc} & = \ln \left ( \f{\muc}{1.5 \! \GeV} \right ) \, , & 
L_{\mub} & = \ln \left ( \f{\mub}{5.0 \! \GeV} \right )
\, ,
\end{aligned}
\end{split}
\eeq
and the sum includes the expansion coefficients $\kappa_{ijkl}$ given
in \Tab{tab:kappas}. The above formula approximates the exact NNLO
result with an accuracy of better than $\pm 1.0 \%$ in the ranges
$1.15 \! \GeV \le \mc (\mc) \le 1.45 \! \GeV$, $0.1150 \le \as (\MZ)
\le 0.1230$, $1.0 \! \GeV \le \muc \le 3.0 \! \GeV$ and $2.5 \! \GeV
\le \mub \le 10.0 \! \GeV$. The uncertainties due to $\mt (\mt)$,
$\muh$ and the different methods of computing $\as (\muc)$ from $\as
(\MZ)$, which are not quantified above, are all below $\pm 0.2
\%$. Their actual size at NNLO will be discussed below.

Using the input parameters listed in \Tab{tab:input}, we find at the
NNLO level    
\beq \label{eq:PcNNLO}
\begin{split}
\Pc & = 0.115 \pm 0.008_{\rm theor} \pm 0.008_{\mc} \pm 0.001_{\as}
\\ 
& = \left ( 0.115 \pm 0.018 \right ) \left ( \f{0.225}{\lambda} \right
)^4 \, ,     
\end{split}
\eeq 
where now the residual scale ambiguities and the uncertainty due to
$\mc (\mc)$ are of the same size. Comparing these numbers with
\Eq{eq:PcNLO} we observe that our NNLO calculation reduces the
theoretical uncertainty by a factor of more than 3.     

As can be nicely seen in the lower plot of \Fig{fig:plots}, $\Pc$
depends very weakly on $\muc$ at NNLO, varying by only $\pm
0.007$. Furthermore, the three different treatments of $\as$ affect
the NNLO result in a negligible way. The three-loop values of $\as
(\muc)$ used in the numerical analysis have been obtained with the
program {\tt RUNDEC}. The theoretical error quoted in \Eq{eq:PcNNLO}
includes also the dependence on $\mub$ and $\muh$ of combined $\pm
0.001$. The presented scale uncertainties correspond to the ranges
given earlier.    

Using \Eqs{eq:BR}, \eq{eq:kappamu}, and \eq{eq:PcNNLO} the result in
\Eq{eq:BRNLO} is modified to the NNLO value 
\begin{align} \label{eq:BRNNLO}
\BR (\Ktomumu)_{\rm SD} & = \left ( 0.79 \pm 0.04_{\Pc} \pm 0.08_{\rm
other} \right ) \times 10^{-9} \non \\
& = \left ( 0.79 \pm 0.12 \right ) \times 10^{-9} \, . 
\end{align}
Obviously, at present the errors from $\mtpole$, $\mc (\mc)$ and the
CKM parameters veil the benefit of the NNLO calculation of $\Pc$
presented in this Letter.  

Provided both $\Pc$ and $\BR (\Ktomumu)_{\rm SD}$ are known with 
sufficient precision useful bounds on the Wolfenstein parameter
$\bar{\rho}$ can be obtained \cite{Buchalla:1993wq}. In particular for
the measured branching ratio $\BR (\Ktomumu)_{\rm SD}$ close to its
SM predictions, one finds that given uncertainties $\sigma (\Pc)$ and
$\sigma (\BR (\Ktomumu)_{\rm SD})$ translate into   
\beq \label{eq:rhobarerror}
\f{\sigma (\bar{\rho})}{\bar{\rho}} = \pm 0.89 \f{\sigma (\Pc)}{\Pc}
\pm 2.59 \f{\sigma (\BR (\Ktomumu)_{\rm SD})}{\BR (\Ktomumu)_{\rm SD}}
\, .
\eeq

As seen in \Eq{eq:rhobarerror} the accuracy of the determination of
$\bar{\rho}$ depends sensitively on the error in $\Pc$. The reduction
of the theoretical error in $\Pc$ from $\pm 22 \%$ down to $\pm 7 \%$
translates into the following uncertainties      
\beq \label{eq:errorcomparison}
\f{\sigma (\bar{\rho})}{\bar{\rho}} = 
\begin{cases}
\pm 20 \% \, , & \hspace{1mm} \text{NLO} \, , \\
\pm 6 \% \, , & \hspace{1mm} \text{NNLO} \, ,  
\end{cases}
\eeq
implying a significant improvement of the NNLO over the NLO
result. In obtaining these numbers we have included only the
theoretical errors quoted in \Eqsand{eq:PcNLO}{eq:PcNNLO}. 

Using the conservative upper bound
\beq \label{eq:BRbound}
\BR (\Ktomumu)_{\rm SD} < 2.5 \times 10^{-9} \, , 
\eeq
on the short-distance part of the $\Ktomumu$ branching ratio derived
in \cite{Isidori:2003ts}, we find the following allowed range 
\beq \label{eq:rhobarbound}
-0.74 < \bar{\rho} < 3.13 \, , 
\eeq
for the Wolfenstein parameter $\bar{\rho}$ employing a customized
version of the \ckm~code.  

To conclude, we have evaluated the complete NNLO correction of the
charm-quark contribution to $\BR (\Ktomumu)_{\rm SD}$. The inclusion
of these contributions leads to a drastic reduction of the theoretical
uncertainty in the relevant parameter $\Pc$. This strengthens the
power of the rare decay $\Ktomumu$ in determining the Wolfenstein
parameter $\bar{\rho}$ from its short-distance branching ratio.   
 
We would like to thank A.~J.~Buras and U.~Nierste for carefully
reading the manuscript, and A.~H\"ocker and J.~Ocariz for useful
correspondence concerning the \ckm~package, and I. Picek and S. Trine
for informative communications regarding the two-loop electroweak
two-photon contribution. U.~H. has been supported by the Schweizer
Nationalfonds. \\

{\bf Note added:} There is an additional contribution from anomalous triangle diagrams to $P_c$ not included in our work. The numerical 
effect of this mistake is negligible, see the erratum of \cite{Buras:2006gb} for details.

\end{document}